\documentclass[final]{aipproc}
\layoutstyle{8x11single}

\usepackage{amsmath,amssymb,amsfonts}
\usepackage{hyperref}
\usepackage{ bbold }

\newcommand{\g}{\mathcal{G}}

\newcommand{\R}{B}

\begin{document}

\title{Astrophysical data analysis with information field theory}


\author{Torsten En{\ss}lin}
{ address = {Max Planck Institut f\"ur Astrophysik (Karl-Schwarzschild-Stra{\ss}e~1, D-85748~Garching, Germany), and
            Ludwig-Maximilians-Universit\"at M\"unchen (Geschwister-Scholl-Platz~1, D-80539~M\"unchen, Germany)}
}

\begin{abstract}
Non-parametric imaging and data analysis in astrophysics and cosmology can be
addressed by information field theory (IFT), a means of Bayesian, data based inference on
spatially distributed signal fields. IFT is a statistical field theory, which permits the construction of
optimal signal recovery algorithms. It exploits spatial correlations of the signal fields even for nonlinear
and non-Gaussian signal inference problems. The alleviation of a perception threshold for
recovering signals of unknown correlation structure by using IFT will be discussed in particular as
well as a novel improvement on instrumental self-calibration schemes. IFT can be applied to many
areas. Here, applications in in cosmology (cosmic microwave background, large-scale
structure) and astrophysics (galactic magnetism, radio interferometry) are presented.
\end{abstract}
\maketitle

\section{Information field theory}
Information field theory (IFT) is information theory for fields, describing in mathematical language how information on spatially distributed quantities following some physical laws can optimally be extracted from data. IFT exploits known or inferred correlation structures of the field of interest $s$ ($s$ is regarded as a fuction $s=s(x)$ and as a vector $s=(s_x)_{x\in \Omega}$ in a function space) over some domain $\Omega = \{ x\}$ in order to regularize the otherwise ill-posed inverse problem of determining virtually infinitely many field degrees of freedom from a finite dataset $d=(d_1,\ldots\, d_\mathtt{n})^\mathrm{T}=(d_i)_i$, $\mathtt{n}\in \mathbb{N}$. What distinguishes it from many non-parametric inference methods is that an IFT is defined over continuous spaces, and any pixelization of the field used in actual computations must preserve this continuum limit and recover it for infinitely small pixels. A concise introduction into IFT can be found  Ref.~\cite{2013AIPC.1553..184E}, exhaustive ones in 
Refs.~\cite{Lemm2003,2009PhRvD..80j5005E}, and the numerical issues of properly discretized fields are addressed in Refs.~\cite{2013A&A...554A..26S,SeligMaxEnt2013}.

Signal field estimation in IFT relies on Bayes theorem that gets recast into a (statistical) field theoretical language:
\begin{equation}
 \mathcal{P}(s|d)=\frac{\mathcal{P}(d|s)\,\mathcal{P}(s)}{\mathcal{P}(d)}\equiv \frac{e^{-\mathcal{H}(d,s)}}{\mathcal{Z}_d}.
\end{equation}
This defines the information Hamiltonian and its partition function to be
\begin{eqnarray}
 \mathcal{H}(d,s)&\equiv& -\ln \mathcal{P}(d,s)= -\ln \mathcal{P}(d|s)-\ln \mathcal{P}(s)\;\mbox{and}\\
 \mathcal{Z}_d &\equiv& \int \mathcal{D}s\, e^{-\mathcal{H}(d,s)} =  \int \mathcal{D}s\, \mathcal{P}(d,s)=\mathcal{P}(d),
\end{eqnarray}
and permits the usage of any appropriate field theoretical technique in order to do our signal inference. These include, among others, classical solutions (or Maximum A Posteriori (MAP) approximation), Feynman diagrams, renormalization and re-summation techniques, Gibbs free energy minimization (or Maximum Entropy), and mean field approaches. See Refs.~\cite{2013AIPC.1553..184E,2009PhRvD..80j5005E} for an overview. IFT can be applied in scientific areas that rely on imaging, like astronomy and astrophysics, and examples of such applications can be found at \href{https://www.mpa-garching.mpg.de/ift}{https://www.mpa-garching.mpg.de/ift}. Useful IFT apporaches to some generic astrophysical signal inference problems are reviewed below.

\section{The measurement problem}
In a generic linear measurement the data depend linearly on the signal,
\begin{equation}
 d=R\,s+n =\left(\int_\Omega \!\! dx\,R_{ix}\,s_x+n_i\right)_i,\label{eq:dRsn}
\end{equation}
with $R$ the instrument response operator, which describes how the instrument senses the signal, and  $n$ the noise. In the simplest case, signal and noise priors are independent Gaussian distributions, 
\begin{equation}
 \mathcal{P}(n,s)= \mathcal{G}(s,\,S)\, \mathcal{G}(n,\,N)\mbox{, with } \mathcal{G}(\phi,\Phi)=\frac{1}{\sqrt{|2\pi \Phi|}}\,\exp\left(-\frac{1}{2}\,\phi^\dagger \Phi^{-1} \phi \right),
\end{equation}
with $S$ and $N$ known signal and noise covariances, respectively, and $\dagger$ denoting the adjoint of a vector or field. Then, the signal posterior is Gaussian as well,
\begin{equation}
 \mathcal{P}(s|d)=\mathcal{G}(s-m,D), \mbox{ with } m=D\,j,\; D=\left(S^{-1} + R^{\dagger} N^{-1}R\right)^{-1},\mbox{ and } j= R^{\dagger} N^{-1}d. \label{eq:WF}
\end{equation}
Here, $m$ is the signal posterior mean or Wiener filter estimate, $D$ the signal uncertainty covariance or information propagator, and $j$ the information source \cite{2009PhRvD..80j5005E}.

A typical signal in astronomy is the intensity distribution $I_x$ of light of a certain wavelength on the celestial sphere $\Omega = \mathcal{S}^2$. Since $I_x$ is strictly positive and can vary over orders of magnitude, the logarithmic brightness $s_x=\log(I_x/I_0)$ would be a natural variable that can be as well positive as negative. The \emph{Maximum Entropy} principle singles out a Gaussian prior for $s$ if only information up to the second moment of the $s$-distribution is available a priori or should be taken into account. Furthermore, the likelihood might be a Poisson distribution, if individual photons are counted. All this complicates the inference problem and introduces so-called interaction terms into the information Hamiltonian, but can in principle be dealt with the toolbox of IFT \cite{2009PhRvD..80j5005E, 2010PhRvE..82e1112E, 2013PhRvE..87c2136O}. 
For simplicity, we assume in the following the measurement Eq.~(\ref{eq:dRsn}).  

In practical applications of IFT, it is often the case that the signal covariance $S$, the noise covariance $N$, the response $R$, or combinations thereof are not sufficiently known for applying the Wiener filter given by Eq.~(\ref{eq:WF}).
We denote all such undetermined parameters with the parameter vector $p$. Its uncertainty has to be taken into account in the signal inference. 
A $p$-marginalized joint probability of data and signal might be calculated analytically,
\begin{equation}
 \mathcal{P}(d,s)=\int\!\!\mathcal{D}p\,\mathcal{P}(d,s,p).
\end{equation}
However, the corresponding information Hamiltonian $\mathcal{H}(d,s) = -\ln \mathcal{P}(d,s)$ is usually a complicated function of the signal, preventing an easy calculation of the posterior mean $m$. Simple MAP signal estimators, resulting from solving $\partial\mathcal{H}(d,s)/\partial s=0$ for $s$, can be highly biased due to the strong skewness of the marginalized posterior \cite{2011PhRvD..83j5014E}. 

The \emph{Empirical Bayes} approach performs much better. There, the $s$-marginalized parameter posterior $\mathcal{P}(d,p)=\int\!\!\mathcal{D}s\,\mathcal{P}(d,s,p)$ is used to choose a good point estimate $p^\star$, which then is assumed to be correct in a signal mean inference based on $\mathcal{P}(s|\,d,p^\star)$.  The reason for \emph{Empirical Bayes} to be a logically consistent approximation can be observed in the following reformulation of the posterior mean signal,
\begin{equation}
 m \equiv \langle s\rangle_{(s,p|d)}\equiv
 \int\!\!\mathcal{D}p\int\!\!\mathcal{D}s\,\mathcal{P}(s,\,p|\,d)\,s 
= \int\!\!\mathcal{D}p\mathcal{P}(p|\,d)\,\int\!\!\mathcal{D}s\,\mathcal{P}(s|\,d,\,p)\,s=\langle\langle s\rangle_{(s|d,p)} \rangle_{(p|d)}\approx \langle s\rangle_{(s|d,p^\star)}.
\end{equation}
In the last step a delta-approximation, $\mathcal{P}(p|d)\approx \delta(p-p^\star)$, was used. Thus, \emph{Empirical Bayes} can be regarded as the zeroth order term of a perturbation expansion in the posterior parameter uncertainty. The next order term might be obtained by using a Gaussian approximation of  $\mathcal{P}(p|d)$. A more systematic expansion can be obtained using the minimal Gibbs free energy approach  \cite{2010PhRvE..82e1112E}. However, already the empirical Bayes approximation leads to well-working signal inference recipes in case of unknown hyper-parameters for the problems of unknown covariances and/or response. 

\begin{figure}[t]
  \resizebox{0.99\textwidth}{!}{\includegraphics{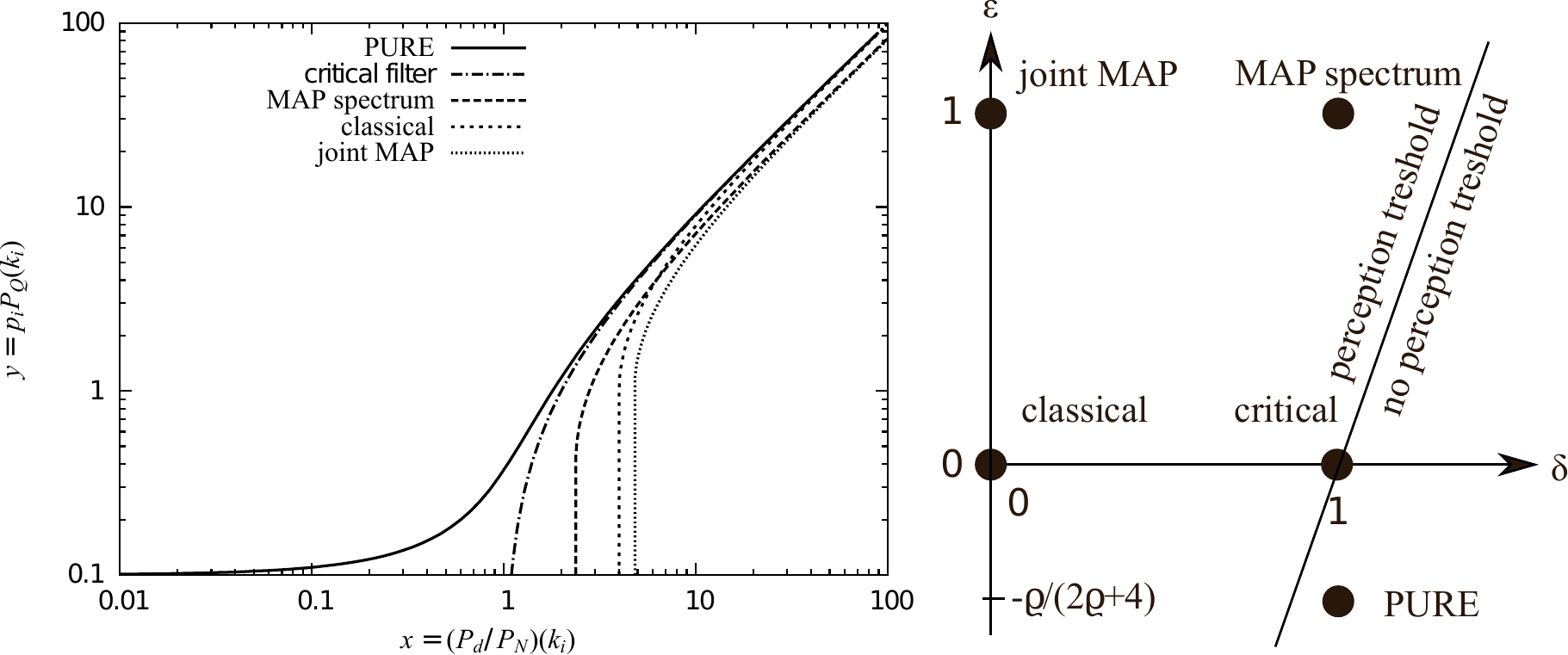}}
%
\label{fig:perception}
\source{Figures from \cite{2011PhRvD..83j5014E}.}
\end{figure}

\section{Unknown signal covariance}
The signal covariance $S$ might be unknown, however, statistical homogeneity can often be assumed, i.e., the correlation $\langle s_x \bar{s_y}\rangle_{(s)}= S_{xy}=C_s(x-y)$ is only a function of the distance between $x$, and $y$. This implies that $S$ is diagonal in Fourier space,
\begin{equation}
 S_{kk'}=(2\pi)^u\,\delta(k-k')\,P_s(k),
\end{equation}
with $k$, $k'$ being $u$-dimensional Fourier wave-vectors. In this case, a suitable hyper-prior for the unknown signal power spectrum $P_s(k)$ might be given by a term penalizing deviations from power-law spectra and inverse Gamma priors for the individual Fourier modes, 
\begin{equation}
 \mathcal{P}(P_s) \propto \exp \underbrace{\left[-\frac{1}{2} \int \!\! dk\,  \left(\frac{\partial \log P_s(k)}{\sigma_k\,\partial \log k}\right)^2\right]}_{\equiv -\frac{1}{2} \log P_s^\dagger T \,\log P_s} \prod_k\, \left[ P_s(k)\right]^{-\alpha_k}\,\exp\left(-\frac{q_k}{P_s(k)}\right). 
\end{equation}
The hyper-parameters $\sigma_k$, $\alpha_k$, and $q_k$ determine how strongly spectral curvature is penalized and how informative the spectral prior is for the individual modes. $\alpha_k=1$ and $q_k=0$ correspond to Jeffreys' non-informative prior.
For the \emph{Empirical Bayes} approximation we have to deduce a point estimate of the spectrum. Using the MAP approximation in $\log P_s(k)$ for this \cite{2013PhRvE..87c2136O}, we obtain
\begin{equation}
 P_s^\star(k) = \frac{q_k+ \frac{1}{2}\,\mathrm{Tr}\left[\left(m^\star\,m^{\star\dagger} + D^\star \right) \mathbb{1}^{(k)} \right] }{\alpha_k -1 +  
 \frac{1}{2} \, \varrho_k+(T\,\log P_s^\star)_k},
\end{equation}
with $\mathbb{1}^{(k)}$ a projection onto all Fourier modes with the same power-spectrum as $k$ (e.g., on spheres in Fourier space, in case of statistical isotropy) and $\varrho_k=\mathrm{Tr}\left[\mathbb{1}_k \right]$ the number of such modes. $m^\star$ and $D^\star$ are the mean and uncertainty covariance, respectively, calculated for the Gaussian inference problem assuming $P_s^\star(k)$ to be self-consistently the correct power spectrum. 

It is instructive to investigate this signal filter operation (the self-consistent calculation of $m^\star$ and $P_s^\star$) in the context of similar filters. For this, we simplify the problem by assuming that signal and noise covariances, as well as response are diagonal in Fourier space, and that our spectral prior is completely non-informative ($\sigma_k=\infty$, $\alpha_k=1$, and $q_k=0$ for all $k$). A generic spectral estimator (to be calculated consistently with the corresponding signal estimator  $m^\star = D^\star j$) is then  
\begin{equation} 
 P_s^\star(k) = \frac{\mathrm{Tr}\left[\left(m^\star\,m^{\star\dagger} + \delta\, D^\star \right) \mathbb{1}^{(k)} \right] }{\varrho_k+\varepsilon},
\end{equation}
where the parameters $\delta$ and $\varepsilon$ have been introduced in order to see how differently signal and spectrum estimators perform. The above derived non-informative filter corresponds to $(\delta,\, \varepsilon)=(1,0)$, but other assumptions or approximations might lead to different values. These are depicted in the top right figure on this page. 
For example, the MAP estimator using the power spectrum marginalized posterior leads to $(\delta,\, \varepsilon)=(0,0)$ (labeled ``classical'') and a parameter uncertainty renormalized estimator (PURE) can be projected to $(\delta,\, \varepsilon)=(1,-\varrho/(2\varrho+4))$. The above estimator with $(\delta,\, \varepsilon)=(1,0)$ is called \emph{critical filter}, because it resides on a critical line for perception thresholds. All the estimators to the left of this line exhibit a perception threshold. They do not respond at all to data modes with power below some critical value. This is depicted in the left figure, where the power in the resulting reconstructions of these filters is shown as a function of the ratio of data to noise power.
Filters with perception threshold require the data to have more variance than expected from the noise. 
For example, the \emph{classical filter} with $(\delta,\, \varepsilon)=(0,0)$ does not account for the missing power in the reconstruction.
As a consequence it has a strong perception threshold. It requires the data to have four times the variance of the noise before it recognizes a signal. The \emph{critical filter} has only a marginal perception threshold at the point where the data variance is exactly the noise variance. The PURE filter tries even to extract information from data with less variance than expected from the noise level.  

\section{Unknown noise covariance}
The critical filter needs to know precisely the noise level of the data in order to tune itself optimally. However, in many measurement situations this is also not reliably known and the data have to provide this information as well. As noise and signal variance are somehow degenerately encoded in the data, an informative prior for the noise level is mandatory whenever a non-informative prior for the signal power spectrum is used.
The noise covariance might be decomposed as $N=\sum_i\,\eta_i\,N_i$, where the $N_i$ denote block matrices in data space that contain the best available guesses for the error covariances. 
We model our noise prior knowledge with a multivariate inverse-gamma prior,
\begin{equation}
 \mathcal{P}(\eta)\propto \prod_i \,\eta_i^{-\beta_i}\,\exp\left( -\frac{r_i}{\eta_i}\right), 
\end{equation}
by choosing $\beta_i>1$ and $r_i>0$ to represent our confidence in the reported or assumed error covariances $N_i$ appropriately. 
The resulting signal estimation scheme is as before, just with an additional estimator for the noise parameters, given by  
\begin{equation}
 \eta_i^\star = \frac{r_i+ \frac{1}{2} \,\mathrm{Tr}\left\{ \left[ (d-R\,m^\star)\,(d-R\,m^\star)^\dagger + R\,D^\star R^\dagger \right] N_i^{-1}\right\}}{\beta_i-1+ \frac{1}{2}\,\mu_i},
\end{equation}
where $N_i^{-1}$ is the pseudo-inverse of $N_i$ and  $\mu_i = \mathrm{Tr}(N_i\,N_i^{-1})$ the dimensionality of the noise blocks \cite{2011PhRvE..84d1118O}. This \emph{extended critical filter} has successfully been used to reconstruct an all sky map of the galactic magnetic field \cite{2012A&A...542A..93O-short}. 

\section{Unknown calibration}

Finally, it might also well be that the response in Eq.~(\ref{eq:dRsn}) is not fully known. The process and the result of determining the response is called \emph{calibration}. Ideally, the measurement of a well-known external calibration signal is used. However, often this is not sufficient, since the instrument response might change with time. 
In this case, the scientific signal of interest might be used as a calibrator itself. The resulting self-calibration scheme (\emph{selfcal}) assumes some calibration, infers the signal with a reconstruction method, assumes this signal to be correct in order to calibrate on this, and iterates until convergence. 
Although the \emph{selfcal} scheme is widely used, in particular in radio astronomy, an information theoretical investigation and a proof about its convergence was only provided recently \cite{2013arXiv1312.1349E}. 
 
There it has been shown that if the imaging and calibration algorithms used can be regarded as MAP estimators, the classical \emph{selfcal} scheme corresponds to a joint MAP estimation of signal and calibration. Similar to the case of unknown signal covariance, we expect such a joint MAP estimation to be sub-optimal in an $\mathcal{L}^2$-error norm sense.

Specifically, we assume the unknown calibration parameters $\gamma=(\gamma_i)_i$ to affect the response linearly,
\begin{equation}
R^{\gamma}=\R^{0}+\sum_{a}\gamma_{a}\R^{a},\label{eq:linearCalibration}
\end{equation}
with $\R^{0}$ and $\R^{a}$ known and $\gamma$-independent, and $\mathcal{P}(\gamma)=\g(\gamma,\,\Gamma)$  the calibration prior with some known calibration uncertainty matrix $\Gamma$.
Using the \emph{Empirical Bayes} approach, we find that the signal reconstruction $m^\star$ should use the calibration point estimate given by Ref.~\cite{2013arXiv1312.1349E} self-consistently 
\begin{eqnarray}\label{eq:sc}
\gamma^{\star} & = & \Delta^{\star}\, h^{\star},\mbox{ with}\\
\begin{gathered}\Delta^{\star}\end{gathered}
_{ab}^{-1} & = & \Gamma_{ab}^{-1}+\mathrm{Tr}\left[\left(m^{\star}\, m^{\star\dagger}+D^{\star}\right)\R^{a\dagger}N^{-1}\R^{b}\right],\mbox{ and}\nonumber \\
h_{b}^{\star} & = & m^{\star\dagger}\R^{b\dagger}N^{-1}d-\mathrm{Tr}\left[\left(m^{\star}\, m^{\star\dagger}+D^{\star}\right)\R^{0\dagger}N^{-1}\R^{b}\right].\nonumber 
\end{eqnarray}
This is a new \emph{selfcal} scheme, which improves over \emph{classical selfcal} by taking  the posterior signal uncertainty $D^\star = \langle (s-m^\star)\,(s-m^\star)^\dagger \rangle_{(s|d,\gamma^\star)}$ into account. 

An example of \emph{selfcal} can be seen in the figure on this page. 
There, an unknown signal is shown, which was scanned three times with an instrument with varying gains (also shown) according to the measurement equation $d_t = (1+\gamma_t)\, s_{x_t}+n_t$ with $x_t= t\,\mathrm{mod} 1$. The solutions of the \emph{classical selfcal} and the \emph{new selfcal} schemes are compared to this. Both solve Eq.~(\ref{eq:WF}) and Eq.~(\ref{eq:sc}) simultaneously, while  \emph{classical selfcal} ignores $D^\star$ in Eq.~(\ref{eq:sc}). The figure shows that the \emph{new selfcal} scheme  alleviates partly a bias of \emph{classical selfcal} for high gain solutions.

\begin{figure}[t]
  \resizebox{0.99\textwidth}{!}{\includegraphics{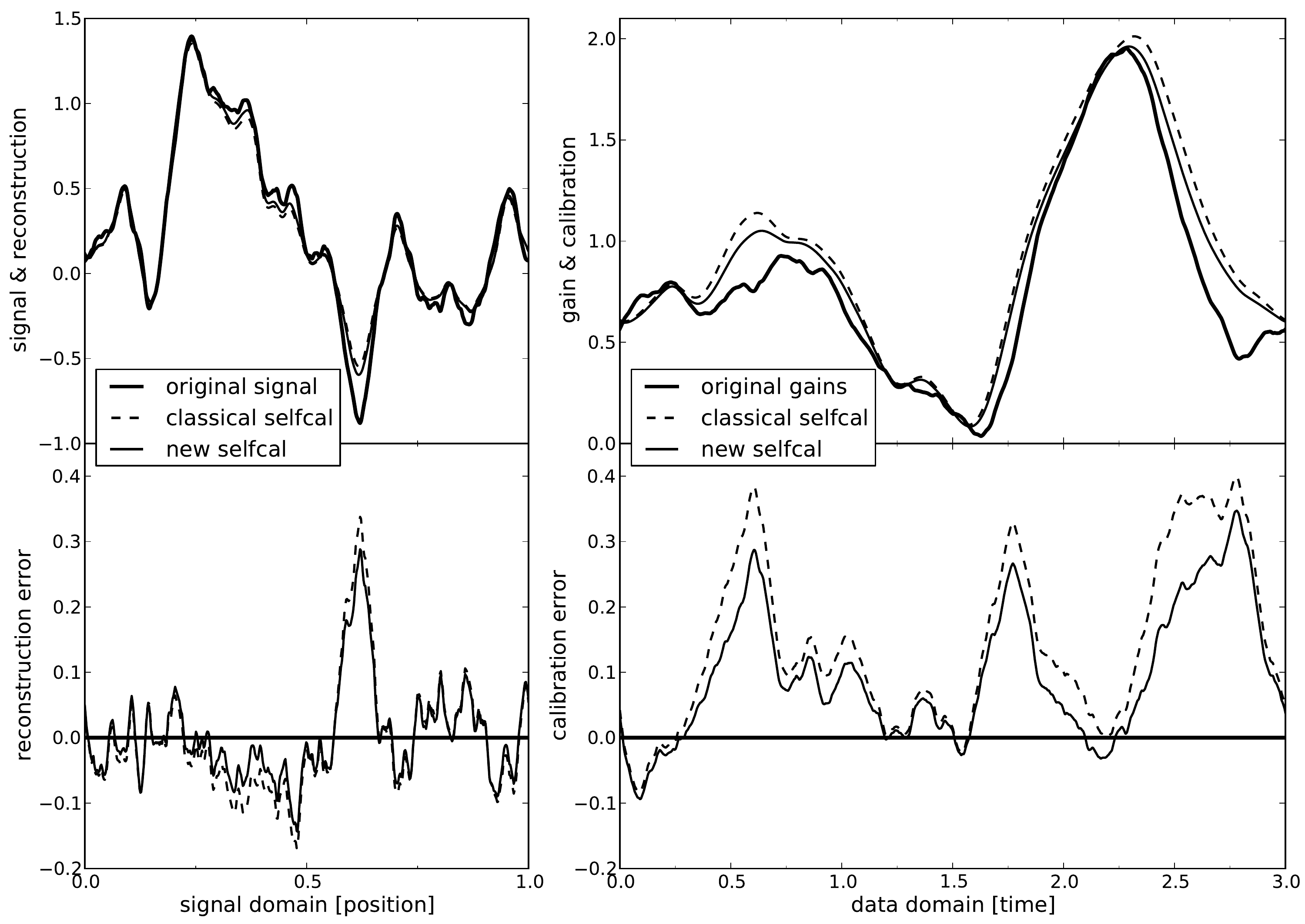}}
%
\label{fig:selfcal}
\source{Figures from \cite{2013arXiv1312.1349E}.}
\end{figure}
\section{Astrophysical applications}

The versatility of IFT to deal with various measurement problems, to take uncertainties on covariances and responses into account and to provide implementable algorithms have let to a number of concrete IFT applications in astronomy and astrophysics. Many of them build on  
\textsc{NIFTy}, Numerical Information Field Theory \cite{2013A&A...554A..26S,SeligMaxEnt2013}, a publicly available, object-oriented library that permits the user to program IFT algorithms abstractly, irrespective of the space discretization used. IFT was already applied to a number of areas, which we briefly discuss in the following.

The \textbf{cosmic microwave background} statistic is nearly Gaussian. However, different inflationary scenarios of cosmology predict different levels of non-Gaussian signatures. Optimal non-Gaussianity estimators could be constructed using Feynman diagrams \cite{2009PhRvD..80j5005E} and by a saddle point approximation \cite{2013PhRvD..88j3516D}.

The \textbf{cosmic large-scale structure} of the matter distribution in the Universe is traced by galaxies and can therefore be reconstructed from galaxy surveys \cite[][and others]{2008MNRAS.389..497K, 2010MNRAS.406...60J, 2012MNRAS.427L..35K}. The power spectrum of the large-scale structure can be measured as well \cite{2010MNRAS.406...60J} and analyzed for its information content on cosmology. Although the cosmic density field is a strictly positive quantity, it is often treated as a Gaussian random field. A log-normal model would be more accurate \cite{2009PhRvD..80j5005E, 2009MNRAS.400..183K-short,  2010MNRAS.409.1393W}. For its usage, it is necessary to translate between linear and logarithmic spectra \cite{2013arXiv1312.1354G}. 

\textbf{Galactic magnetism} can be studied using the Faraday rotation of extragalactic radio sources. The construction of all-sky Faraday maps from such individual measurements towards the directions of these sources was possible thanks to the \emph{critical filter} \cite{2011A&A...530A..89O} and the \emph{extended critical filter} \cite{2012A&A...542A..93O}. 

\textbf{Interferometric radio astronomy} faces the problem that an interferometer array measures only some part of the Fourier transformed sky brightness. The other part has to be reconstructed by the imaging algorithm. \emph{RESOLVE}, a novel IFT based imaging method, which assumes the sky brightness to follow a log-normal distribution with unknown power spectrum, seems to be superior in reconstructing diffuse emission compared to classical algorithms (e.g. \emph{CLEAN} and \emph{MAXENT}) \cite{2013arXiv1311.5282J}.

\textbf{Gamma- and X-ray astronomy} have to deal with extended emission, point sources, Poisson statistics, inhomogeneous sky exposure, and complex point spread functions. The IFT based D$^3$PO algorithm for \emph{denoising, deconvolving, and decomposition of photon observations} handles these and produces maps of the extended emission, catalogs of the point sources, and angular power spectra of the diffuse flux \cite{2013arXiv1311.1888S}.


These applications demonstrate that IFT is a versatile framework to develop and analyze measurement and imaging problems not only in astronomy, but also in cosmology and other areas.

\begin{theacknowledgments}

First of all, I should apologize for exclusively reviewing IFT work I was involved in. There is a large amount of related work I had to ignore for brevity. I want to thank my IFT co-investigators of the here discussed works, Michael Bell, Sebastian Dorn, Mona Frommert, Maksim Greiner, Jens Jasche, Henrik Junklewitz, Righi Khatri, Francisco Shu Kitaura, Niels Oppermann, Martin Reinecke, Georg Robbers, Marco Selig, and Cornelius Weig. Furthermore, I acknowledge helpful comments on the manuscript by Vanessa B{\"o}hm, Sebastian Dorn, and Marco Selig.


\end{theacknowledgments}

\newcommand{\aap}{A\&A}
\newcommand{\apj}{APJ}
\newcommand{\jcp}{JCP}
\newcommand{\mnras}{MNRAS}
\newcommand{\prd}{Phys.Rev.D}
\newcommand{\pre}{Phys.Rev.E}

\bibliographystyle{aipproc}

\bibliography{../../../Text/Bib/ift}

\end{document}